\documentclass{article}
\usepackage{graphicx} 
\usepackage{subfigure}
\usepackage{cite}
\usepackage{amsmath}
\usepackage{amssymb}
\usepackage{subfig}
\usepackage{float}
\usepackage{geometry}
\usepackage{authblk}

\title{Discovery of knowledge of wall-bounded turbulence via symbolic regression}
\author[a]{ZhongXin Yang}
\author[a]{XiangLin Shan}
\author[a]{WeiWei Zhang}
\affil[a]{Northwestern Polytechnical University, Xi’an 710072, China}
\date{May 2024}

\begin{document}

\maketitle

\begin{abstract}
With the development of high performance computer and experimental technology, the study of turbulence has accumulated a large number of high fidelity data. However, few general turbulence knowledge has been found from the data. So we use the symbolic regression (SR) method to find a new mixing length formula which is generally valid in wall-bounded turbulence, and this formula has physical interpretation that it has correct asymptotic relationships in viscous sublayer,buffer layer, log-law region and outer region. Coupled with Reynolds averaged Navier-Stokes (RANS) solver, we test several classic cases. The prediction results fully demonstrate the accuracy and generalization of the formula. So far, we have found that SR method can help us find general laws from complex turbulent systems, and it is expected that through this 'white box' machine learning method, more turbulence knowledge with physical interpretation can be found in the future.
\end{abstract}

\section{Introduction}
From ancient to modern, the main ways to conduct scientific research have gradually changed. First, Humanity's earliest scientific research, characterized primarily by the recording and description of natural phenomena, is called experimental science(the first paradigm), such as the famous Reynolds experiment.
However, due to the limitations of experimental conditions, it is difficult to achieve a more accurate understanding of nature, so theoretical science (the second paradigm) is emerged which through theoretical simplification, leaves only the key factors for research. A successful example is the Reynolds averaged Navier-Stokes (RANS) equation\cite{reynolds1895iv}.
Later, with the birth of electronic computers, it became possible to solve partial differential equations (PDEs), so numerical simulation (the third paradigm) was born. During this period, the development speed of human beings was significantly improved, and many disciplines were derived, such as computational fluid mechanics\cite{anderson1995computational}.\
With the further development of computer, the data is expected to explode from 33 zettabytes in 2018 to 175 zettabytes in 2025. Then data-intensive sciences\cite{hey2009fourth} comes into public view. This approach to science is known as the fourth paradigm.
With the help of  Artificial Intelligence (AI), a lot of research in fluid dynamics has been conducted.
In the aspect of flow control,Yu Zhou et al \cite{RN179}proposed an artificial intelligence control system to maximize the mixing rate of turbulent jets.
In terms of mode reduction,Murata et al.\cite{RN183} proposed a nonlinear mode decomposition method for flow field based on convolutional neural networks, which has good feature separation potential compared with Proper Orthogonal Decomposition method.
In the area of turbulence modeling ,Dehao Xu et al. proposed a nonlinear algebraic model based on neural networks for large eddy simulation (LES) of compressible wall turbulence\cite{RN181}. Park et al. proposed a neural network-based SubGrid-Scale model (SGS) for incompressible channel turbulent flow \cite{RN182}. 

Many of the above are examples of using data for fluid dynamics research, and the fundamental logic of using big data analysis for scientific research is that the generation of data necessarily contains knowledge\cite{RN184}. The function of AI is to replace people to mine this law in massive data. There is already a lot of work combining AI with fluid mechanics. Julia Ling et al.\cite{RN50} proposed a neural network that satisfies Galilean invariance and obtained better results than naive neural networks without embedded invariance. Zhu et al.\cite{RN152} and Shan et al.\cite{RN155}directly constructed a turbulence model using deep neural networks and achieved good results on airfoil flow. But the shortcoming is that this method is difficult to have no reasonably physical explanation. Peng Cui et al.\cite{RN187}pointed out that physical interpretability is a key problem to be solved in machine learning at present. Due to the high complexity of many application areas, machine learning models struggle to explain algorithmic processes and predictive results to humans. However, in areas such as health care, financial justice, etc., the potential risk of incorrect predictions due to the lack of physical explanation can bring great harm and is an urgent problem in the field of machine learning. The symbolic regression has the characteristic of good physical interpretation. Symbolic regression works by finding an analytic parameterized function that best describes the given data\cite{RN185}. It has been applied to materials science\cite{RN158}, astrophysics \cite{RN107,RN108}. And it has been gradually applied in the field of fluid mechanics. Jun Zhang et al.\cite{RN186} used the sparse recognition method to obtain the correct form of the governing equation from the data simulated by the molecule and accurately determine the value of the transport coefficient. Wu Chengyu et al.\cite{RN149} modified the generation term coefficient of the SST turbulence model by using the symbolic regression method to enhance the prediction accuracy of the SST model in separated flows. The essence of the success of these works is that the development of things follows the objective physical laws, and the knowledge is hidden in the observation data. How to discover knowledge from the data has become a new research paradigm!

In this paper, we used SR method for the first time to mine a physically interpretable mixing-length formula from high-fidelity data, and calculate several cases coupled with RANS solver. Therefore, the accuracy and generalization of the mixing-length formula are fully tested. The prediction results show that the formula is more accurate, compared with the traditional turbulence model in practical calculation.

The rest of the paper is organised as follows. The symbolic regression procedure and formula discoverd are described in Part 2. The interpretability of the formula is given in Part 3. The assessment of the formula under the RANS framework is shown in Part 4. Finally , conclusions are given in Part 5.

\section{Discovery of a new Mixing-Length formula}
Despite its simplicity,Prandtl's mixing-length hypothesis\cite{RN164} is still a cornerstone of our understanding of wall-bounded turbulence. Prandtl says that l l 'may be considered as the diameter of the masses of fluid moving as a whole in each individual case;or again,as the distance traversed by a mass of the type before it becomes blended in with neighbouring masses....';and also that l is 'somewhat similar,as regards effect,to the mean free path in the kinetic theory of gases'.Pirozzoli\cite{RN29} also suggested close connection between the mixing length and the physical size of the outer-layer eddies.

Under the boussinesq eddy hypothesis,the Reynolds stress and mean velocity gradient have a relationship as follows:
\begin{equation}
    -\overline{{u}^{\prime}{v}^{\prime}} = {\nu }_{T}\frac{\partial \overline{u}}{\partial y}
\end{equation}
where $-\overline{{u}^{\prime}{v}^{\prime}}$ is Reynolds stress,${\nu }_{T}$ is eddy viscosity and $\frac{\partial \overline{u}}{\partial y}$ is mean velocity gradient.To describe the eddy viscosity ,a mixing length,$l$,is introduced here.
\begin{equation}
    {\nu }_{T}={l}^{2}\left | \frac{\partial \overline{u}}{\partial y}\right |
\end{equation}

In any turbulence boundary layer flow,the mixing length can be computed from measured profiles of velocity and shear stress.
\begin{equation}
    {l}^{2}=\frac{-\overline{{u}^{\prime}{v}^{\prime}}}{\partial \overline{u}/\partial y }
    \label{ml cal}
\end{equation}

Next, symbolic regression (SR) method will be introduced. Like fully connected neural network ,SR is also a machine learning method.But different from the former,it can get an explicit mathematical expression , which can be explained.So it's called 'White box' model ,while neural network is named 'black box'.In this paper,We use a symbolic regression method from a third party python library called PySR\cite{cranmer2023interpretable}.

In this section, the different physical factors (e.g., intermittent,pressure gradient) are broken down into multiple steps for regression to ensure that the effects of different physical factors will not be confused by machine learning, and also to help improve the accuracy of the formula. In the first step, the basic mixing length relationship will be discovered from the fully developed channel flow; The second step is to find the outer region characteristics of mixing length according to the intermittent characteristics of turbulence in boundary layer flow. In the third step, considering the influence of adverse pressure gradient, the mixing length formula is modified by pressure gradient correction parameter(\textbf{$\gamma$}).This strategy is called the three-step strategy

\subsection{Step1}

\begin{figure}[ht]
    \centering
    \includegraphics[width=0.5\linewidth]{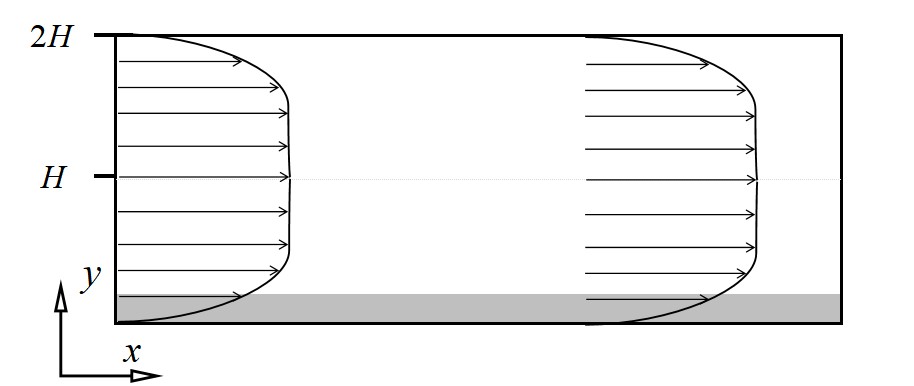}
    \caption{Channel flow}
    \label{fig:ch}
\end{figure}

Fully developed channel flow is shown in Fig.\ref{fig:ch}.The simulations reported here are DNS of incompressible turbulent flow between two parallel planes. Periodic boundary conditions are applied in the streamwise (x) and spanwise (z)directions, and no-slip/no-penetration boundary conditions are applied at the wall.The computational domain sizes are $L_x = 8\pi H$ and $L_z = 3\pi H$, where H is the channel half-width, so the domain size in the wall-normal (y) direction is 2H. The flow is driven
by a uniform pressure gradient.

The DNS was carried out by Myoungkyu Lee et al.\cite{lee2015direct}
And in step 1, the DNS data will be used to compute the mixing length,according to  Eq.(\ref{ml cal}).The channel data of the grey part at two friction Reynolds numbers $Re_\tau$($Re_\tau=\frac{H u_\tau }{\nu}$)are used as training set, while the data of the other three friction Reynolds numbers will be used to test the generalization ability of the formula, as shown in Table 1.

\begin{table}[h!]
\centering
\begin{tabular}{c|c}
\hline
\textbf{$Re_\tau$} & \textbf{Property} \\
\hline
180  & test     \\ \hline
550  & test     \\ \hline
1000 & train    \\  \hline
2000 & test     \\  \hline
5200 & train    \\  \hline
\end{tabular}
\caption{Channel flow data}
\end{table}

In the first step,five dimensionless variables are given as input variables in Eq.(\ref{step1:input}).
\begin{equation}
    x_0=\frac{y}{H}, x_1=Re_\tau,x2={y}^{+},x_3=\sqrt{\frac{y}{H}},x_4=\sqrt{{y}^{+}}
    \label{step1:input}
\end{equation}
where $y^+=\frac{\sqrt{\rho\tau_w}}{\mu}y$.To reduce the difficulty of symbolic regression, the Van Driest'damping function(Eq.(\ref{VD}))\cite{van1956turbulent} is be used to get a higher fidelity mixing length formula.

\begin{equation}
    l=\kappa y(1-{e}^{-\frac{{y}^{+}}{{A}^{+}}})
    \label{VD}
\end{equation}

The objective function is listed in Eq.(\ref{targ1}).

\begin{equation}
    y_{obj,1}=\frac{l_{hf,1}}{\kappa y(1-{e}^{-\frac{{y}^{+}}{{A}^{+}}})}
    \label{targ1}
\end{equation}
where $\kappa = 0.41$,and subscript obj,1 and hf,1 represent the objective function  and the mixing length computed from high fidelity(hf) data by Eq.(\ref{ml cal}) in step1  respectively.

Table.\ref{operator} shows operators used in symbolic regression.

\begin{table}[h!]
    \centering
    \begin{tabular}{c|c}
    \hline
    Operator Type     & Operators  \\ \hline
    Unary operators   & $exp(x_j),ln(x_j),tanh(x_j)$ \\ \hline
    Binary operators  & $x_i+x_j,x_i*x_j,\frac{x_i}{x_j},1-exp(-\frac{x_i}{x_J})$\\ \hline
    \end{tabular}
    \caption{Operator used in SR}
    \label{operator}
\end{table}

Loss function is defined in Eq.(\ref{loss})
\begin{equation}
    Loss_{u}(E) = ln(abs(\frac{y_{predict} - y_{target}}{y_{target}})+1)
    \label{loss}
\end{equation}
E denotes a generated expression by PySR,$y_{predict},y_{target}$ represent the values predicted by E and expected values respectively.Subscript u means loss function is defined by users.In addition to the user-defined loss function, PySR adds a complexity penalty mechanism, due to the traditional philosophy of scientists, the simpler the expression is the easier to interpret by scientists. Thus, PySR allows adaptive adjustment of penalty complexity to expect that the number of expressions of different complexity are roughly equal. PySR calculates the complexity of each expression by Eq.(\ref{complexity}).
\begin{equation}
    C(E) = N_{var} + N_{const} + N_{op}
    \label{complexity}
\end{equation}
where $N_{var},N_{const},N_{op}$ denote the complexity of variables,constant and operator respectively.C(E) represents the complexity of the expression E.
PySR computes the complexity of each expression by Eq.(\ref{compx overall}) finally.
\begin{equation}
    Loss(E) =Loss_u(E)*exp(frecency[C(E)])
    \label{compx overall}
\end{equation}
where $frecency[C(E)]$ represents the ratio of expressions which have the same complexity with expression E.

Finally,combine the most reasonable expression chosen from generated expressions with with Eq.(\ref{VD}).The mixing length discoverd is listed in Eq.(\ref{l1})
\begin{equation}
    l=\kappa y\frac{(1-{e}^{-\frac{{y}^{+}}{{A}^{+}}})}{1-e^{-\frac{\sqrt{y^+}}{3.6}} + 1.21\frac{y^+ +\sqrt{y^+}}{e^{\sqrt{y^+}}+8.19}}e^{-\frac{y}{H}}
    \label{l1}
\end{equation}
After observation , Eq.(\ref{l1}) cab be simplified as follows.
\begin{equation}
    l=\kappa y f_{inner}(y^+) g_{outer}(\frac{y}{H})
    \label{simplify ml}
\end{equation}
where $f_{inner}(y^+)$ is:
\begin{equation}
    f_{inner}(y^+) = \frac{(1-{e}^{-\frac{{y}^{+}}{{A}^{+}}})}{1-e^{-\frac{\sqrt{y^+}}{3.6}} + 1.21\frac{y^+ +\sqrt{y^+}}{e^{\sqrt{y^+}}+8.19}}
    \label{finner}
\end{equation}
$g_{outer}(\frac{y}{H})$ is:
\begin{equation}
    g_{puter}(\frac{y}{H}) = e^{-\frac{y}{H}}
    \label{gouter}
\end{equation}
The subscript 'inner' represents that the input of Eq.(\ref{finner}) is inner scale $y^+$,and its value is approximately 1 in the outer layer ($y^+$ \textgreater 50).On the contrary,the subscript 'outer' denotes the input of Eq.(\ref{gouter}) is outer scale $\frac{y}{H}$ , and it does not affect the inner layer( $\frac{y}{H}$ \textless 0.1).Because its value is approximately 1 in the inner layer. 

\subsection{Step2}

Considering that the main flow condition in the aviation is external flow, such as flow around airfoil. unlike channel flow, there is an intermittent flow phenomenon. The actual outer edge of the turbulent boundary layer is an extremely irregular and unsteady interface. At the  position near the nominal outer edge of the boundary layer ($\delta_{99}$), the flow is turbulent for sometime and laminar for another time. This phenomenon of alternating turbulence and laminar flow at the same point in space is called intermittent phenomenon. Klebanoff \cite{klebanoff1955characteristics} tested the distribution of the intermittent factor in a turbulent boundary layer along a smooth plate, and found that the interval mainly appears outside 0.4 times the thickness of the boundary layer. Therefore, once again we consider the mixed length formula of the external flow, considering the intermittenness,the outer layer function should be replaced  . The new outer layer function is found from the turbulent boundary layer of a flat plate with zero pressure gradient, and the data is derived from the large eddy simulation data of Georg Eitel-Amor et al.\cite{eitel2014simulation}. The simulation was carried out on a flat plate, and the transition was triggered by the tripping device. The dimensions of the physical domain in the
streamwise, wall-normal, and spanwise direction are $L_x\times L_y\times L_z = 13500\times 400\times540$ measured in displacement thickness at the inlet $\delta^*$ ,whereas an open boundary is modelled in the wall-normal direction. Since the purpose of this step is to fine the  outer layer function of the mixing length in external flow, only the data located at the outer layer($0.2\delta_{99} ~0.8\delta_{99}$ exactly) is selected as the training set, and a total of 11 sections which the friction Reynolds number ranging from 450 to 2500 are used.
Below are input variables in step 2.

\begin{equation}
    x_0 = \frac{y}{\delta_{99}},x_1=Re_\tau
    \label{step2:input}
\end{equation}
The objective function is shown in Eq.(\ref{targ2}).

\begin{equation}
    y_{obj,2}=\frac{l_{hf,2}}{\kappa y f_{inner}(y^+)}
    \label{targ2}
\end{equation}
where $f_{inner}(y^+)$ is Eq.(\ref{finner}).Besides ,operators and loss function are the same in step1.
Through this step,another outer layer function is discovered .
\begin{equation}
    h_{outer}=tanh(l_k(\frac{y}{\delta_{99}})^{-1})
    \label{houter}
\end{equation}
where $\delta_{99}$ is the thickness of boundary layer ,and $l_k$ is a function of $Re_\tau$.Its form is shown here.
\begin{equation}
    l_k = 0.215 + (Re_\tau + 1.0086^{Re_\tau})
    \label{lk}
\end{equation}

\subsection{Step3}
In addition to the intermittent of boundary layer flow, the pressure gradient caused by curvature effects is also not negligible in industrial applications (e.g., wing surface curvature).Pressure gradients are ubiquitous in real flows, and their effects are elusive. At present, there is no specific theoretical framework for the boundary layer of pressure gradient to describe the influence of pressure gradient\cite{bobke2017history, gungor2016scaling}.To understand the boundary layer flow under pressure gradient ,Townsend\cite{townsend1956properties, townsend1961equilibrium} proposed a definition of equilibrium boundary layer,but it is extremely strict. But Townsend and Mellor$\&$Gibson\cite{mellor1966equilibrium} proposed another similar definition named near-equilibrium boundary layer and how to satisfy near-equilibrium conditions.Near-equilibrium boundary layer should satisfy the following relation:
\begin{equation}
    U_\infty = C(x-x_0)^m
    \label{Uinf_near}
\end{equation}
Where C and m are constants , $x_0$ is the virtual origin.Townsend pointed out m should be greater than $-\frac{1}{3}$ to achieve near-equilibrium condition.Besides,Clauser\cite{clauser1956turbulent} summarized a parameter to describe pressure gradient $\beta = \frac{\delta^+}{\tau_w} \frac{dp_e}{dx}$ ($\delta^*$ is the thickness of boundary layer,$\tau_w$ is the wall shear stress,$p_e$ is the pressure at the outer edge of the boundary layer) to quantitatively characterize the magnitude of pressure gradient.

Data used here is simulated by well resolved large eddy simulation under different adverse pressure gradient.Details can be found in ref\cite{RN122}.Listed below are data sets for training.Here ,only inner layer where ($y^+ <50$) are used.A pressure gradient force first influences the inner layer and then the outer layer. Additionally, the pressure gradient initially affects the mean flow, through which it influences the turbulence. Therefore, in terms of turbulence modeling, the inner layer necessitates more significant pressure gradient corrections than the outer layer. This is why we concentrate on calibrating the model within the inner layer where $y^+$ is less than 50.

\begin{table}[h!]
    \centering
    \begin{tabular}{c|c|c|c}
    \hline
       Case  &  $Re_\theta$ & $\beta$     & m      \\   \hline
        M16  &  [1010,4000] & [1.55,2.55] & -0.16  \\   \hline
        B1   &  [910,3360]  & $\approx 1$ & -0.14  \\   \hline
    \end{tabular}
    \caption{Adverse Pressure Gradient flat plate boundary layer flow data}
    \label{APG data}
\end{table}
$Re_\theta$ is Reynolds number range based on momentum thickness.

The input variables is listed below.
\begin{equation}
    x_0 = y^+,x_1=P^+
    \label{step3：input}
\end{equation}
where $P^+=\frac{\nu}{\rho {u_\tau}^3} \frac{dp_e}{dx}$ is dimensionless pressure gradient.
The target funtion is :
\begin{equation}
    y_{obj,3}=\frac{l_{hf,3}}{\kappa y f_{inner}(y^+) h_{outer}(\frac{y}{\delta_{99}},Re_\tau)}
    \label{targ3}
\end{equation}
Other setting is the same as Step1.

Finally,we discovered the function and called it by the name of $\gamma$
\begin{equation}
    \gamma = 1 + 18 P^+(1-tanh(\frac{y^+-12.7}{14}))
    \label{gamma}
\end{equation}

\section{The interpretability of the formula}
We summarize Eq.(\ref{simplify ml}),Eq.(\ref{gouter}),Eq.(\ref{houter}),Eq.(\ref{gamma}) here.
\begin{equation}
    l = \kappa y f_{inner}(y^+)F_{outer}\gamma
    \label{sumarized ml1}
\end{equation}
\begin{equation}
    F_{outer}=\begin{cases}
        e^{-\frac{y}{H}}, for\thinspace channel\thinspace flow\\
        tanh(l_k(\frac{y}{\delta_{99}})^{-1}),for\thinspace boundary\thinspace layer\thinspace flow
              \end{cases}
    \label{Fouter}
\end{equation}
where $f_{inner}$ is Eq.(\ref{finner}),and $\gamma$ is Eq.(\ref{gamma}),and $l_k$ is Eq.(\ref{lk}).

A similar form was first proposed by Cebeci and Bradshaw\cite{cebeci2013physical}. They combined Eq.(\cite{van1956turbulent}) with Eq.(\ref{nikuradse}) resulting in Eq.(\ref{cb ml}),and found it valid for the whole pipe flow. 
\begin{equation}
    \frac{l}{R} = 0.14 - 0.08(1-\frac{y}{R})^2 -0.06(1-\frac{y}{R})^4
    \label{nikuradse}
\end{equation}

\begin{equation}
    \frac{l}{R} = (0.14 - 0.08(1-\frac{y}{R})^2 -0.06(1-\frac{y}{R})^4)(1-{e}^{-\frac{{y}^{+}}{{A}^{+}}})
    \label{cb ml}
\end{equation}
Later , You Wu  et al.\cite{wu2013karman} used a similar form in their SED-SL model. These examples can prove that the law truly exists in the data, whereas it was previously deduced through theory, but for the first time we have mined this law directly from the data.

\begin{figure}[ht]
        \centering
        \includegraphics[width=0.5\linewidth]{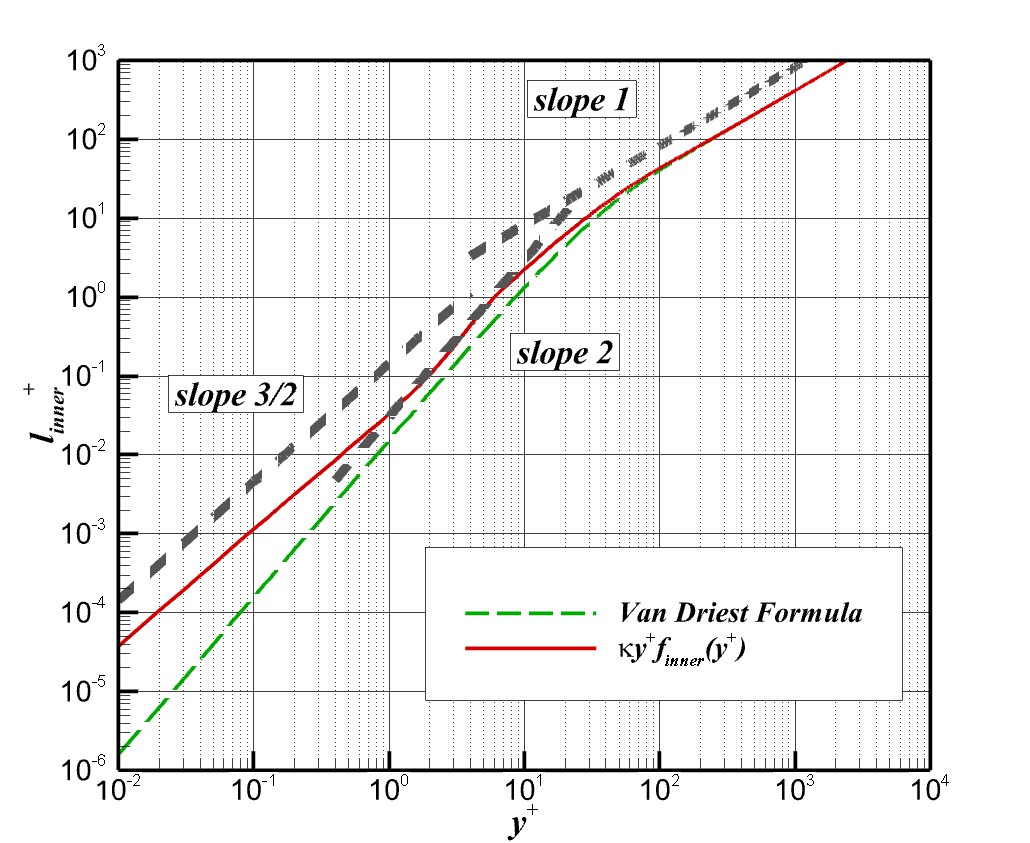}
        \caption{Asymptotic analysis chart}
        \label{fig:finner mlsr}
\end{figure}
For a better explanation of Eq.(\ref{sumarized ml1}) and Eq.(\ref{Fouter}),two functions will be introduced to analyze their mathematical characteristics.
\begin{equation}
    {l_{inner}}^+ = \kappa y^+ f_{inner}(y^+)
    \label{linner}
\end{equation}
\begin{equation}
    l_{outer} = \kappa y F_{outer}
    \label{louter}
\end{equation}
In Fig.\ref{fig:finner mlsr},it can be seen that there are three simple relationships,${l_{inner}}^+ \backsim {y^+}^\frac{3}{2}$ in viscous sublayer,${l_{inner}}^+ \backsim {y^+}^2 $ in buffer layer, and ${l_{inner}}^+ \backsim {y^+}^1$ in low-law region respectively.${l_{inner}}^+ \backsim {y^+}^\frac{3}{2}$ can be obtained by asymptotic analysis near the wall. The explaination of ${l_{inner}}^+ \backsim {y^+}^2 $ can be referred to the work of Xi Chen et al.\cite{chen2018quantifying}. ${l_{inner}}^+ \backsim {y^+}^1$ is the most commonly used law of the wall(Low).

\begin{figure}[htbp]
    \centering
    \begin{minipage}{0.49\linewidth} 
            \centering
            \includegraphics[width=0.9\linewidth]{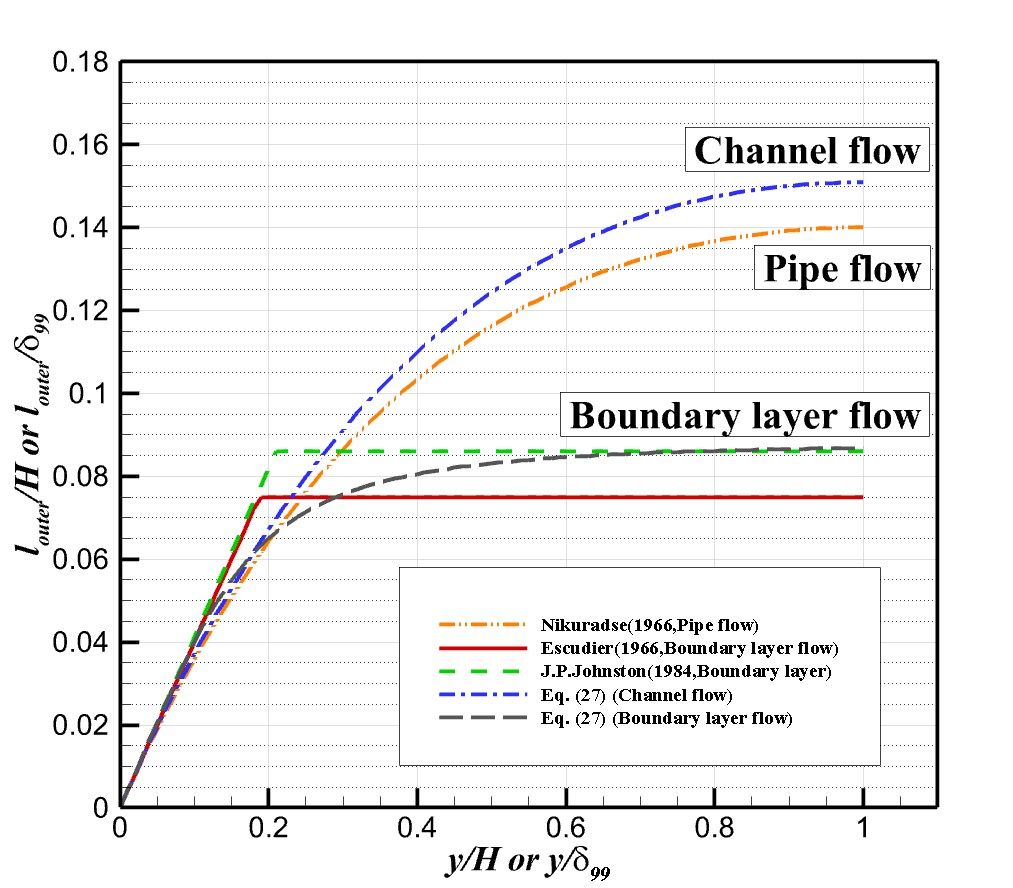}
            \caption{$l_{outer}$ in diffenrent flow}
            \label{fig:louter}
    \end{minipage}
    \begin{minipage}{0.49\linewidth}
            \centering
            \includegraphics[width=0.9\linewidth]{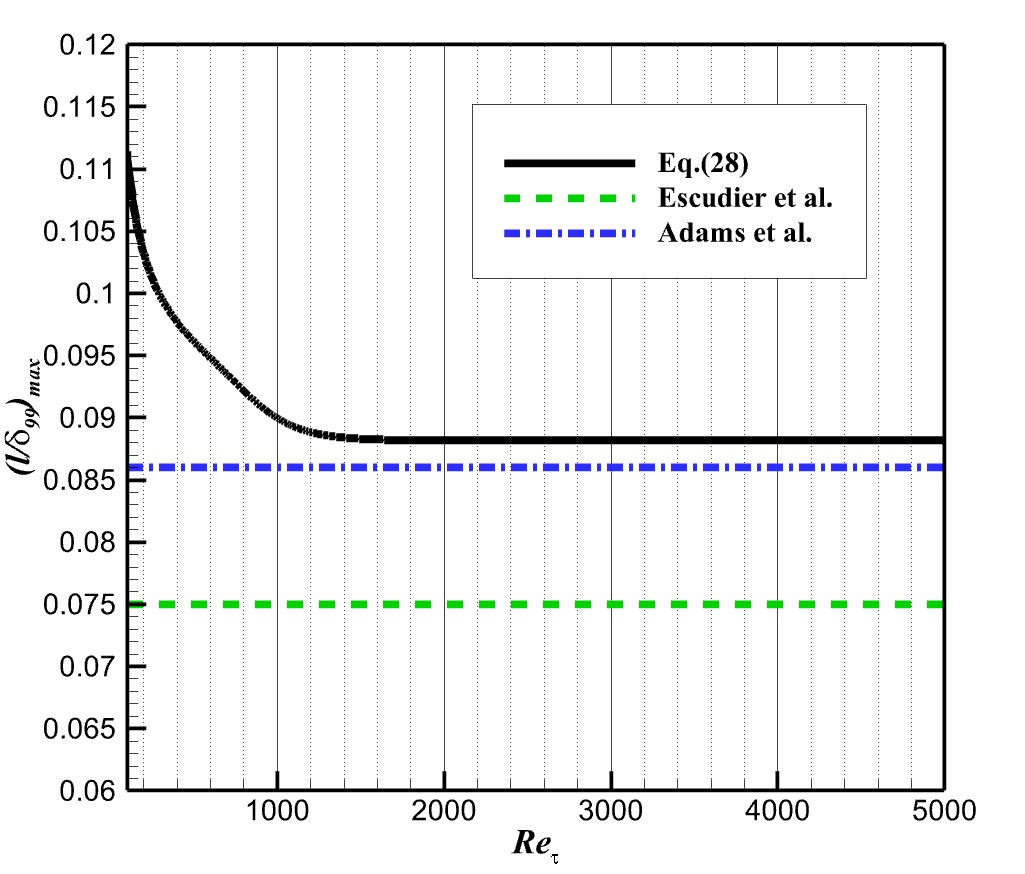}
            \caption{the maximum value of mixing length in Boundary layer flow}
            \label{fig:max}
    \end{minipage}
\end{figure}

Next we show the outer region properties of the mixing length in Fig.\ref{fig:louter}, where the value of $l_k$ in Eq.(\ref{Fouter}) as a function of $Re_\tau$ is taken to be 0.215.And 0.215, as we will explain later in Fig.\ref{fig:max}, is also an specially asymptotic value at high Reynolds number.

Escudier et al.\cite{escudier1966note} and Adams et al.\cite{adams1984mixing} both gave the maximum values for the mixing length in boundary layer flow , 0.075 and 0.086 times the boundary layer thickness, respectively, and it is found that Adams's results have significantly higher accuracy by our modern symbolic regression approach . For the channel flow, the mixing length in the outer region is significantly larger than for the boundary layer flow due to the absence of intermittent, which again seems to be in agreement with the current literature\cite{monty2007large}, where the outer-layer eddies are found to be smaller in the boundary layers and larger in channels, based on Prandtl's physical interpretation of the mixing length.
In Fig.\ref{fig:louter},it is known that the mixing length monotonically increases within the boundary layer, and so the mixing length takes a maximum value at the edge of boundary layers.
\begin{equation}
    (\frac{l}{\delta_{99}})_{max}=\kappa \frac{\delta_{99}}{\delta_{99}} tanh(l_k(\frac{\delta_{99}}{\delta_{99}})^{-1}) \approx \kappa l_k
    \label{eq.max}    
\end{equation}
So , 0.215 is the maximum value of the mixing length at high Reynolds number divided by the Karmen constant.

 Plotted in Fig.\ref{fig:max},Eq.(\ref{eq.max}) gives the trend of the maximum values of mixing length at low Reynolds numbers, and in the next part, it can be found that because of capturing the law of change of mixing lengths at low Reynolds numbers, the simple formula plays an excellent role in numerical simulation with almost the same results as the direct numerical simulation(DNS).

\section{Assessment and analysis}
We have carried out the physical analysis of the mixing length formala, but how it can be used for the simulation of complex flows to construct a highly accurate turbulence model for engineering becomes significant.

At present, since the flow condition in engineering is basically high Reynolds number flow, the RANS method is still the main method In the RANS framework,the unclosed quantity-Reynolds stresses, are modeled using the turbulence model.So the predictive results are the best indicator of the turbulence model.

\begin{figure}
    \centering
    \includegraphics[width=0.8\linewidth]{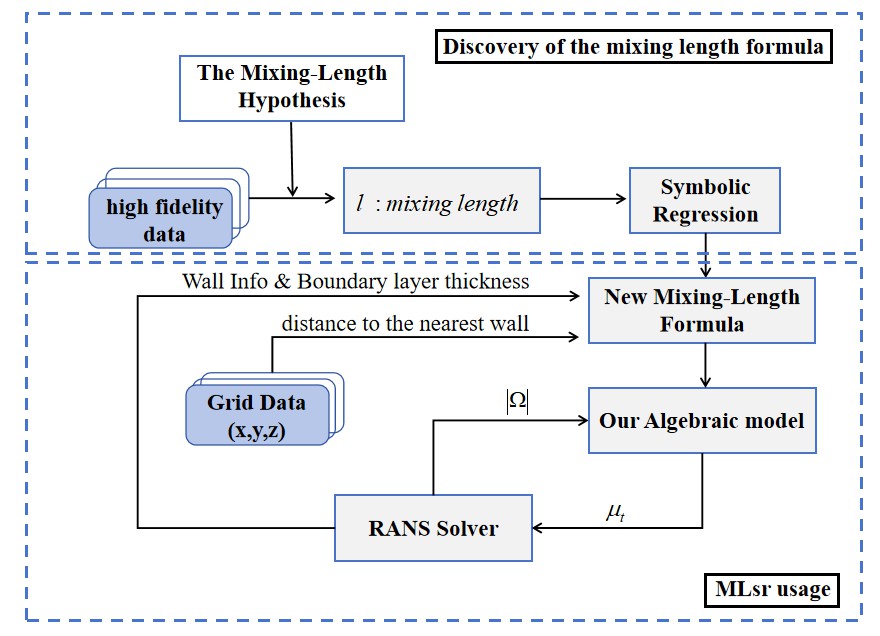}
    \caption{Discovery and usage}
    \label{fig:DIS and use}
\end{figure}

In the calculation process,We use the same boundary layer thickness determination method as Baldwin et al.\cite{baldwin1978thin}
\begin{equation}
    \delta_{99} = \frac{1}{0.65} arg max(l\left | \Omega \right |)
    \label{bl deter}
\end{equation}
where,$\left | \Omega \right |$ is the vorticity magnitude. And the eddy viscosity can be calculated by Eq.(\ref{eddy viscosity}).
\begin{equation}
    \nu_t = l^2\left | \Omega \right |
    \label{eddy viscosity}
\end{equation}
This model is called Mixing Length symbolic regression(MLsr),Summarized in Fig.\ref{fig:DIS and use}.

\subsection{Case1:Fully developed Channel(FDC)}
\begin{figure}[h!]
    \begin{minipage}{0.49\linewidth}
        \centering
        \includegraphics[width=0.9\linewidth]{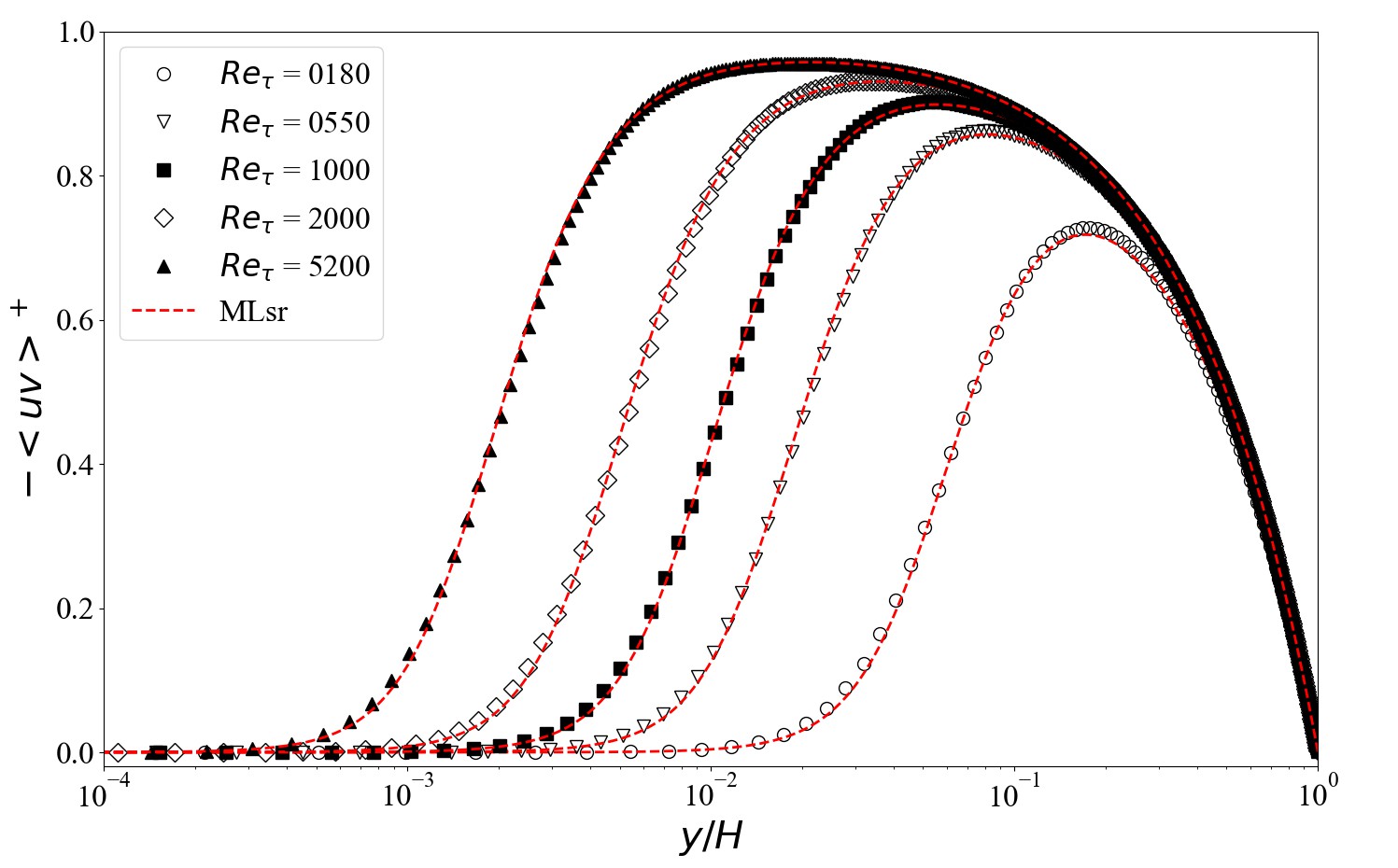}
        \caption{Reynolds stresses}
        \label{fig:rs}
    \end{minipage}
    \begin{minipage}{0.49\linewidth}
        \centering
        \includegraphics[width=0.9\linewidth]{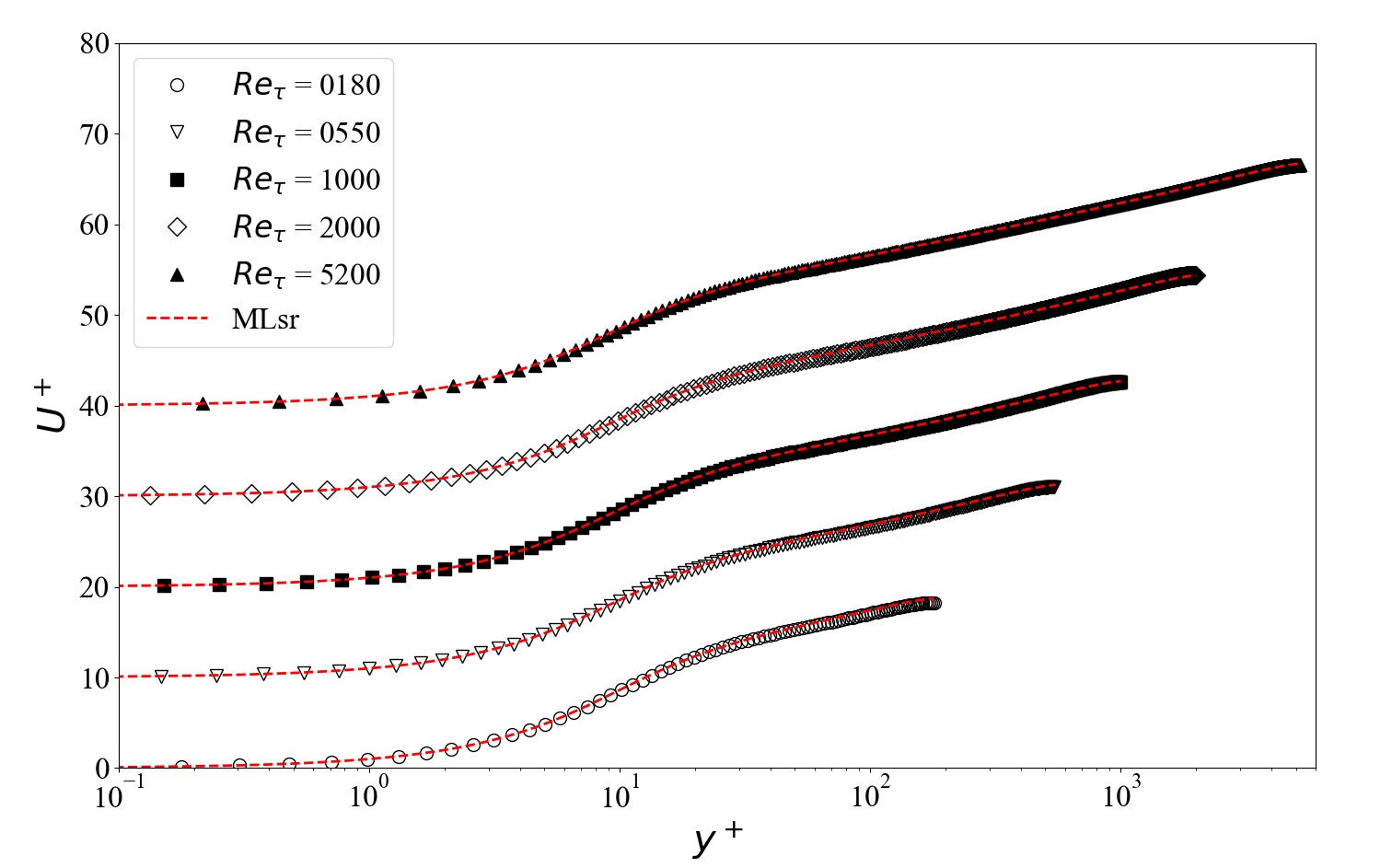}
        \caption{Velocity profiles}
        \label{fig:vp}
    \end{minipage}
\end{figure}

The fully developed channel flow is described  in Step . We use MLsr to predict the remaining three sets of flow condition($Re_\tau = 180,550,2000$) 
The predicted Reynolds stresses and velocity profiles are shown in Fig.\ref{fig:rs} and Fig.\ref{fig:vp}.
The prediction results show that both the velocity profiles and the Reynolds stresses predicted by MLsr are highly close to the DNS data.Based on the execellent results,we draw a conclusion that SR method mines the real physical laws from data and results prove its validation in the low friction Reynolds number flow ,even though the data used for training is only a portion of the flow field data of two high friction Reynolds numbers (1000,5200). This example strongly proves that mining physics formulas from data is not nonsense. And after the discovery of the effective physical formulas, it provides a theoretical approach for the further theoretical study of the flow mechanism of wall turbulence.

\subsection{Case2:Zero Pressure Gradient Flat Plate Boundary Layer(ZPGFPBL)}
This example is incompressible flow ,and the  boundary conditions are said in Step2

\begin{figure}[htb]
    \centering
    \includegraphics[width=0.49\linewidth]{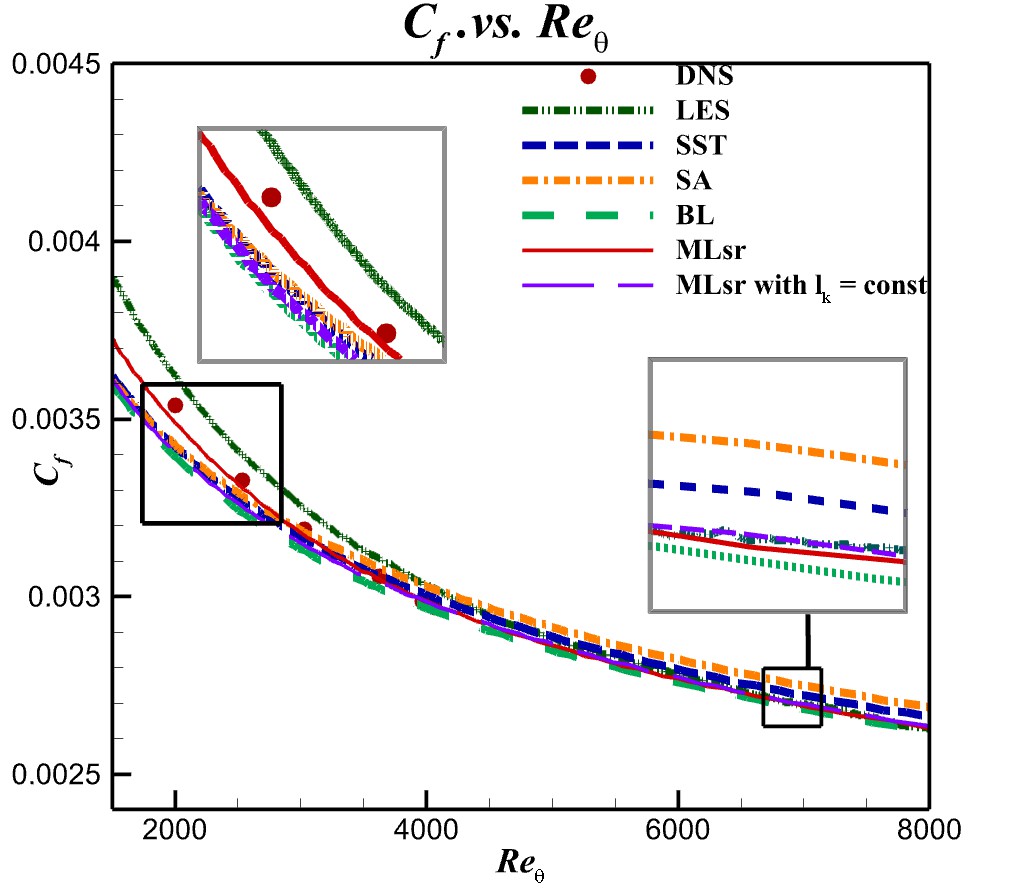}
    \caption{$C_f - Re_\theta$ in ZPGFPBL}
    \label{fig:ZPG cf}
\end{figure}

Fig.\ref{fig:ZPG cf} plots the surface friction coefficient as a function of $Re_\theta$,which is the momentum loss thickness based Reynolds number. By observing Fig.\ref{fig:ZPG cf}, it is obvious that at low momentum thickness Reynolds number, the DNS\cite{schlatter2010assessment}, LES and RANS data all show deviations, but DNS is the most accurate numerical simulation method without no models involved. Surprisingly, the results predicted by MLsr are highly close to the DNS results at low momentum thickness Reynolds number.If we set $l_k = 0.
215$ then the MLsr model without considering the low Reynolds number effect predicts the same results as SST and SA at low Reynolds number, which also reflects that the low Reynolds number effect does not seem to be taken into account in the modeling of these two models.

\begin{figure*}[h!]
	\centering
	\subfigure[$Re_\theta \approx 8200$]{
		\begin{minipage}[t]{0.45\linewidth}
			\centering
			\includegraphics[width=2.5in]{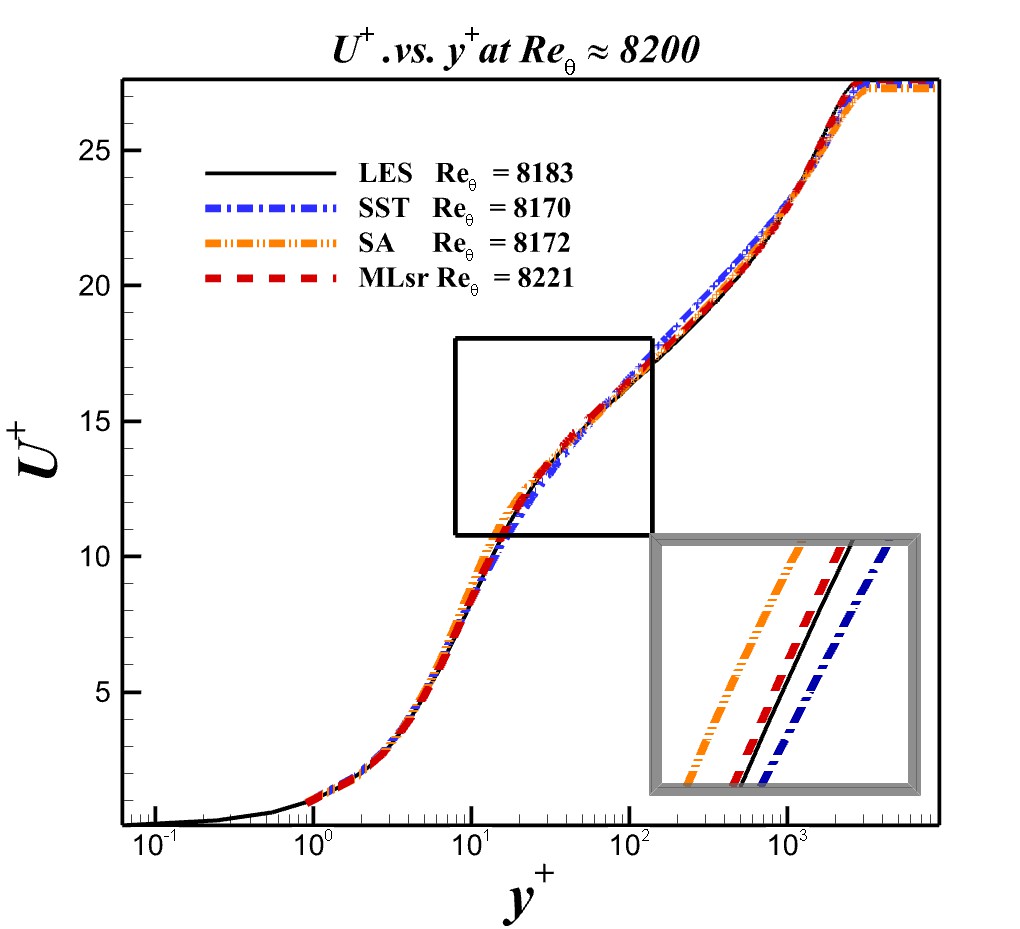}\\
			\vspace{0.02cm}
			\includegraphics[width=2.5in]{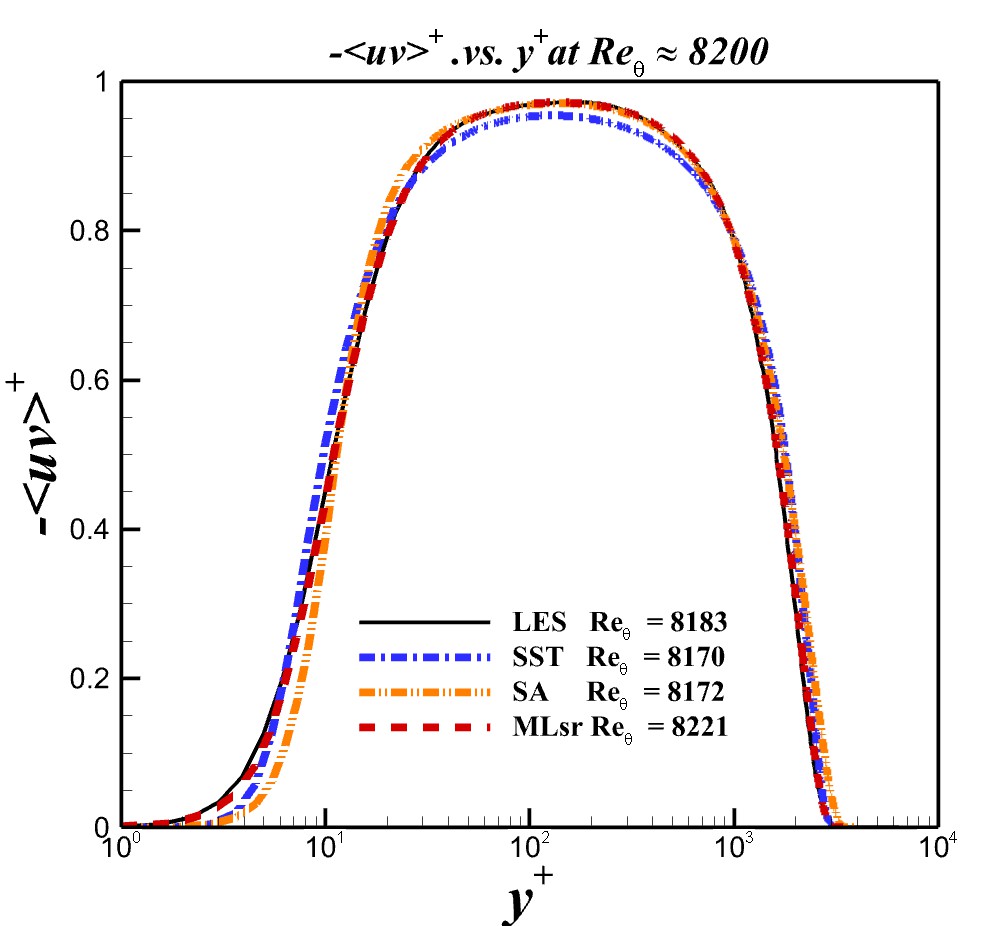}\\
		\end{minipage}%
	}%
	\subfigure[$Re_\theta \approx 2000$]{
		\begin{minipage}[t]{0.48\linewidth}
			\centering
			\includegraphics[width=2.5in]{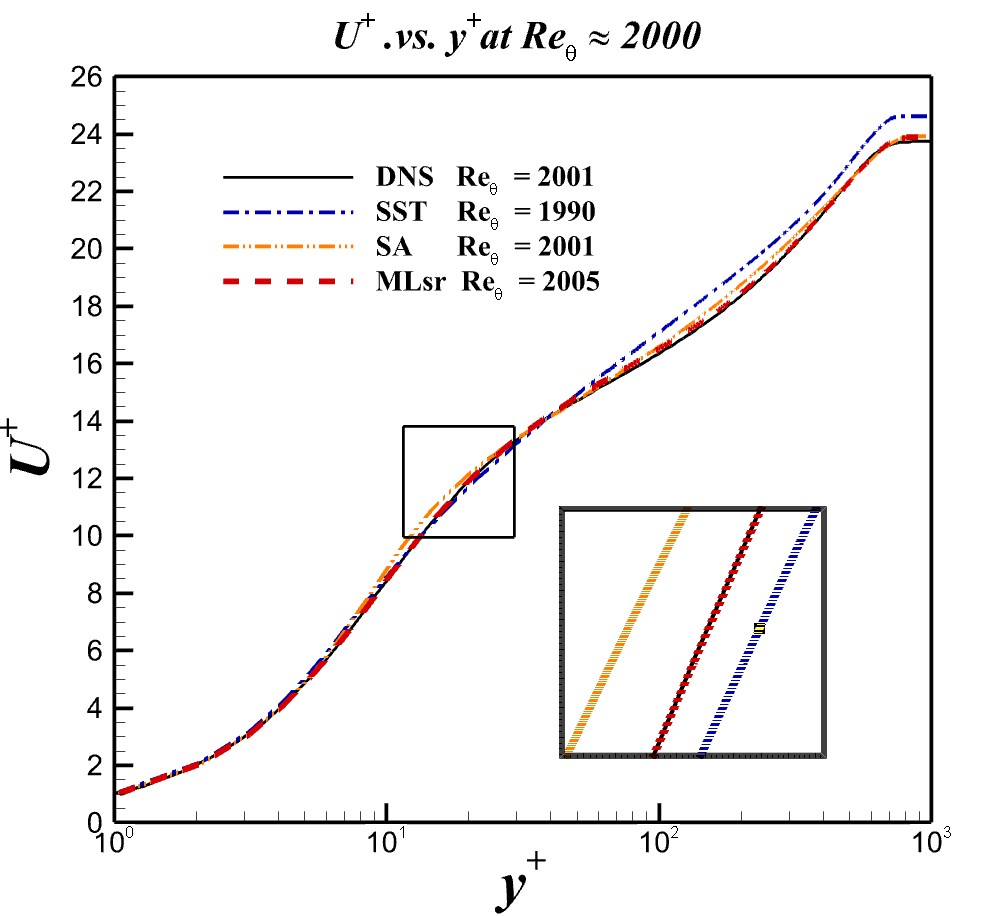}\\
			\vspace{0.02cm}
			\includegraphics[width=2.5in]{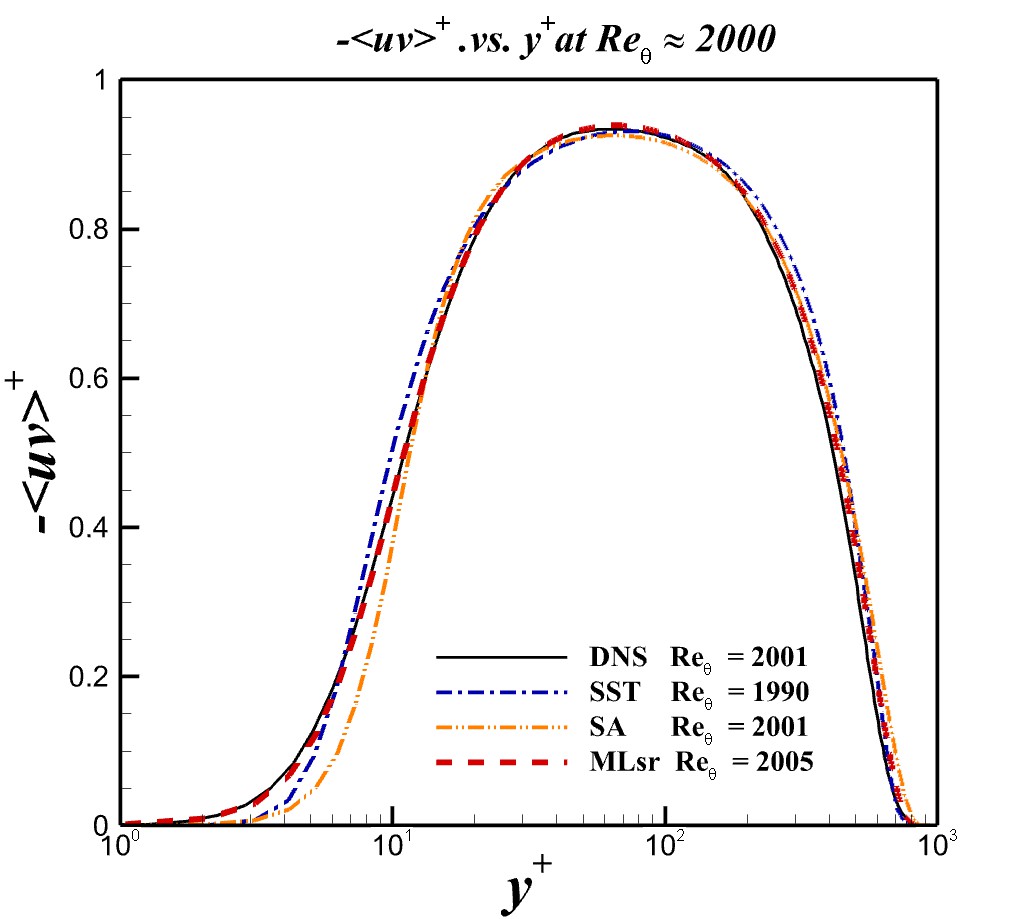}\\
			\vspace{0.02cm}
		\end{minipage}%
	}%

	\centering
	\caption{Two sections in ZPGFPBL}
	\vspace{-0.2cm}
	\label{section}
\end{figure*}

From Fig.\ref{ZPG fig}  , it can be seen that MLsr model outperforms the currently commonly used SST and SA models in both velocity profiles and Reynolds stress distributions. In particular, In viscous sublayer and buffer layer, MLsr model predicts significantly better Reynolds stress distributions than other turbulence models.It is not only because MLsr model is mined from high fidelity data, but also because the correct asymptotic relationship is found, which is commonly established in wall turbulence. Moreover, as can be seen, at low momentum thickness Reynolds number, the prediction results of MLsr are highly consistent with DNS results.
Through the prediction results of the flat plate boundary layer flows, both at high momentum  thickness Reynolds number and low momentum thickness Reynolds number, MLsr is much better than the existing turbulence models, and the inner function of mixing length formula is learned from channel flows, and the outer function is adapted according to the boundary conditions of the flow, but the prediction results show that the inner function is found to be adapted to two different types of wall-bounded flows, reflecting the consistent effect of the wall on the wall-bounded turbulence. We are once again firmly convinced that the mixing length formulas learned from the data captures the true physics of wall-bounded turbulence.

\subsection{Case3:Imcompressible flow over airfoil}
Airfoils are the basic elements in the design of airplane wings, tail fins, missile airfoils, helicopter rotors, propellers, and wind turbine blades, which directly affect the aerodynamic performance of aerospace vehicles, so the prediction of airfoil flow is extremely crucial. Next, the famous subsonic airfoil NACA0012(Fig) will be used to verify the prediction capability and accuracy of MLsr.

\begin{figure}[htb]
    \centering
    \includegraphics[width=0.49\linewidth]{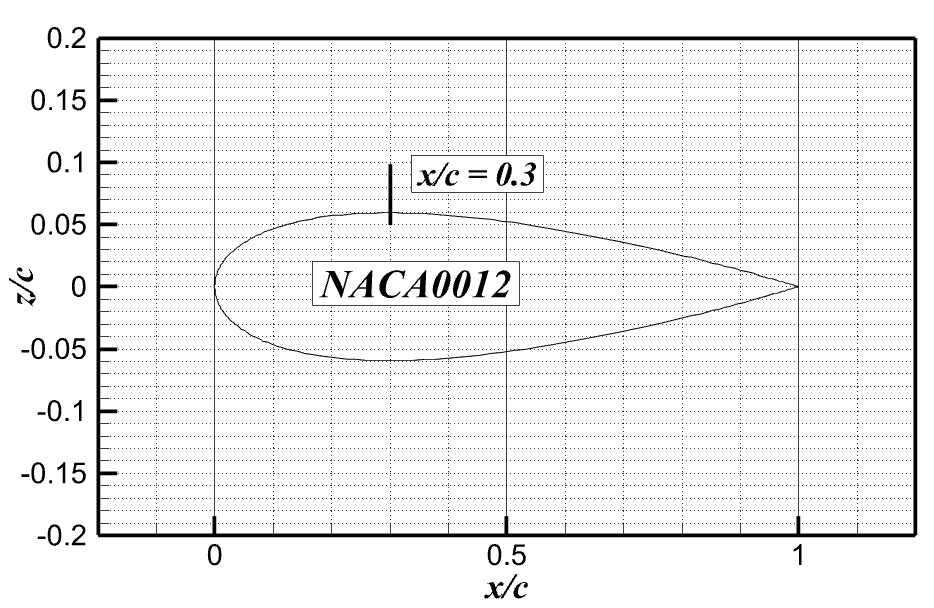}
    \caption{NACA0012}
    \label{fig:nc12}
\end{figure}

The flow field of the NACA0012 airfoil(see Fig.\ref{fig:nc12}) with a Reynolds number of 400,000 is computed at an angle of attack of 0 degrees using different  turbulence models, and the results predicted by the different models were compared with data of the well-resolved LES\cite{tanarro2020effect}. This large eddy simulation accurately resolves the large-scale turbulence, while the smallest scales are modeled with the SubGrid stress model\cite{schlatter2004transitional}.
\begin{figure}[h!]
    \begin{minipage}{0.49\linewidth}
        \centering
        \includegraphics[width=0.9\linewidth]{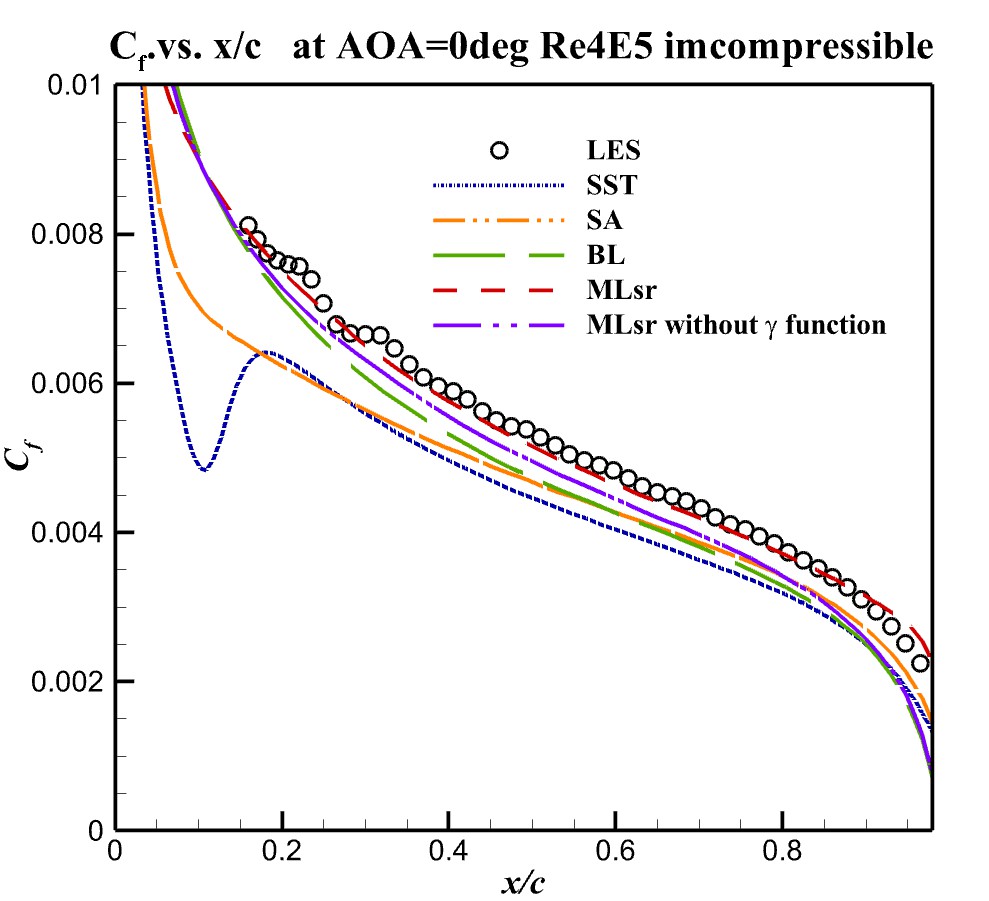}
        \caption{$C_f - x/c$}
        \label{fig:na cf}
    \end{minipage}
    \begin{minipage}{0.49\linewidth}
        \centering
        \includegraphics[width=0.9\linewidth]{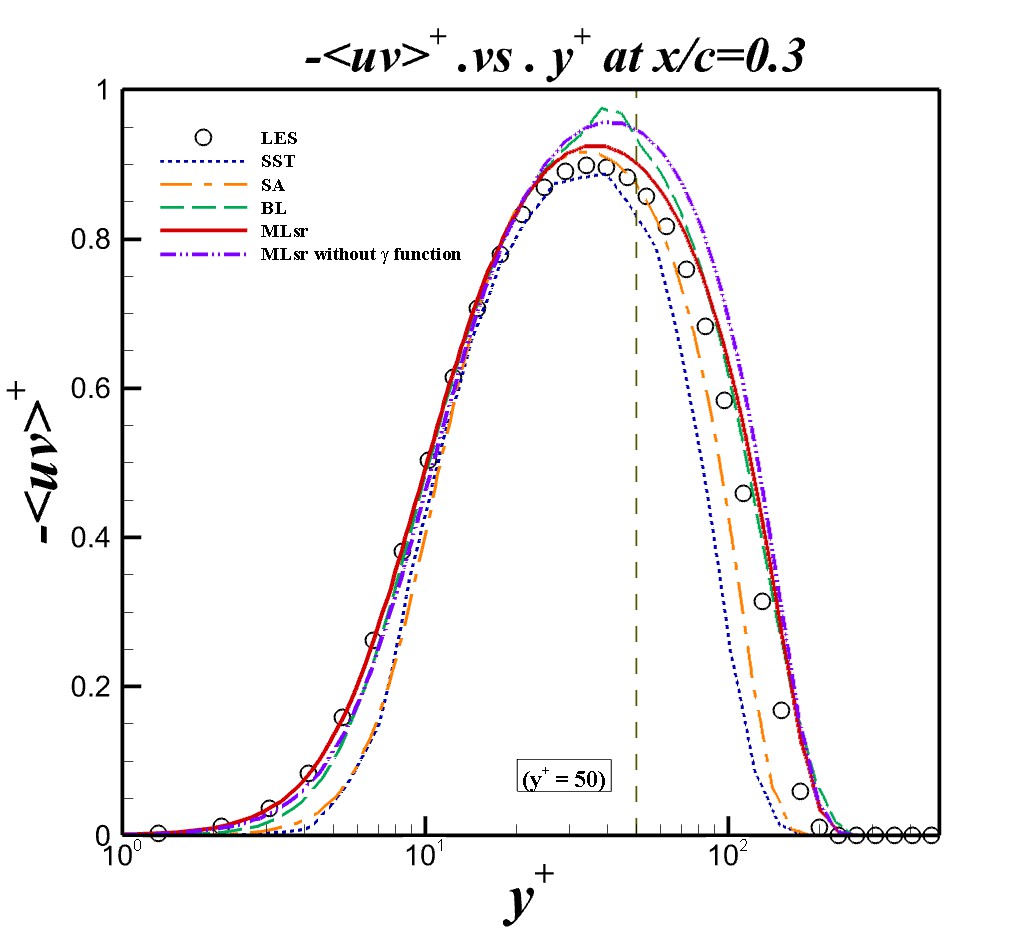}
        \caption{$-<uv>^+ - y^+(x/c = 0.3)$}
        \label{fig:uv 0.3}
    \end{minipage}
\end{figure}
In Fig.\ref{fig:na cf}, MLsr predicts the surface friction stress distribution closest to the LES data, which is better than other models. Moreover, in this low Reynolds number case, the SST model and SA model results have large errors with the LES results, which seems to prove again the conclusion obtained from Case2 that these two models are not applicable to low Reynolds number flows. Meanwhile, we also performed the simulations before and after the removal of the $\gamma$function as a correction for the pressure gradient, and as we expected, the results deteriorated to some extent after the removal. In Fig.\ref{fig:uv 0.3}, it is found that the pressure gradient correction only performed in the range ($y^+ < 50$) is able to trend consistently closer to the LES results for the whole cross section.

\section{Conclusions}
In the field of AI4Science research, symbolic AI has the potential to make a resurgence as a mainstream approach alongside connectionism. Symbolic AI, which focuses on the manipulation of symbols and rules to represent knowledge and perform reasoning tasks, offers unique advantages such as interpretability, transparency, and explicit representation of domain knowledge. By leveraging symbolic AI techniques in scientific applications, researchers may be able to better understand and explain the processes and decisions made by AI systems, leading to more reliable and trustworthy results. With the increasing demand for explainable AI in scientific research, symbolic AI could play a key role in unlocking new insights and advancing the state of the art in AI4Science.In this research ,We have made useful attempts in the field of turbulence.

In this paper, we mine mixing length formulas from DNS and LES data based on the mixing length hypothesis. In the process of mining physical formulas using symbolic regression, a three-step strategy is used creatively to fully mine high fidelity mixing length formulas .First, the basic forms of the mixing length are discovered directly from the data of the channel flow. For boundary layer flow, the effect of intermittency on the mixing length is taken into account, so we again learn the outer function for boundary layers flow from the data of the flat plate boundary layer. Thirdly,in addition to the generality of the existence of pressure gradients, their effects are also elusive.  We thus learned pressure gradient corrections only at locations($y^+ < 50$) where this effectively avoids the complexities of history effects.Coupled with RANS solver ,the mixing length formula is used to predict channel flow, flat plate boundary layer flow, and airfoil flow. The results fully demonstrates the prediction accuracy of the formula. In the results of NACA0012, we are surprised to find that the prediction of $C_f$ is  really close to the LES results after the pressure gradient correction is applied to the mixing length , which is a strong example to tell us that that the influence of complex factors such as history effects seems to have little effect on the prediction of surface friction stresses. The formula is simple enough that it is very easy to embed into the CFD code. However, simplicity is also a disadvantage in that it does not reflect the non-local nature of turbulence, and in the future we can try to bring it directly into the framework of wall-modelled large eddy simulations to serve as a high-precision wall model.

\bibliographystyle{apalike}

\bibliography{main}

\end{document}